\documentclass[12pt]{iopart}

\usepackage{graphicx}
\begin{document}

\title{The pseudogap behaviors in FeSe superconductor ($T_c\sim$9.4 K)}

\author{Yoo Jang Song$^{1}$, Jong Beom Hong$^{1}$, Byeong Hun Min$^{1}$, Kyu Jun Lee$^{2}$, Myung Hwa Jung$^{2}$, Jong-Soo Rhyee$^{3}$, and Yong Seung Kwon$^{1,*}$ }\footnote{$^{*}$Author to whom any correspondence should be addressed.}

\address{$^1$ Department of Physics, Sungkyunkwan University,
Suwon 440-746, Republic of Korea\\$^2$ Department of Physics, Sogang University, Seoul 121-742, Republic of Korea\\$^3$ Samsung Advanced Institute of Technology, Yongin 446-712, Republic of Korea}
\ead{yskwon@skku.ac.kr}
\begin{abstract}
This paper reports the synthesis and superconducting behaviors of the tetragonal iron-chalcogenide
superconductor FeSe. The electrical resistivity and magnetic moment
measurements confirmed its superconductivity with a $T_c^{zero}$
and $T_c^{mag}$ at 9.4 K under ambient pressure. EPMA indicated the sample to have a stoichiometric Fe:Se ratio of 1:1 ($\pm$0.02).
The Seebeck coefficient which was 12.3 $\mu$V/K at room temperature,
changed to a negative value near 200 K, indicating it to be a two carriers material.
Above $T_c$, the $\rho(T)$ curve revealed an 'S' shape. Hence
$d\rho(T)/dT$, and $d^2\rho(T)/dT^2$ showed pseudogap-like behavior
at $T^*$=110 K according to the resistivity curvature mapping (RCM) method
for high $T_c$ cuprates. Moreover, the magnetoresistance $\rho_H(T)/\rho_{H=0}$ under a magnetic field  and
the Seebeck coefficient $S(T)$ revealed revealed pseudogap-like behavior near $T^*$.
Interestingly, at the same temperature, 30 K, the sign of $S(T)$ and all signs of $d^2\rho(T)/dT^2$ 
changed from negative to positive above $T_c$.

\end{abstract}

\vspace{2pc}
\noindent{\it Keywords}:
\maketitle

\section{Introduction}
~~~~Since the discovery of the LaFeAsO$_{1-x}$F$_x$ superconductor, various types
of iron-based superconductors with layered structures containing FeP, FeAs, FeSe,
and FeTe layers have been reported. Hsu \textit{et al.}~\cite{Hsu08} reported superconductivity for FeSe$_{1-x}$
with $T_c$=8 K, at $x$=0.12 and 0.18. Since then, the $T_c$ of the iron-selenide superconductor has been increased rapidly to 27 K at pressure $P$=1.48 GPa~\cite{Mizuguchi08}. Recent studies of the pressure effect on
iron-selenide~\cite{Sidorov09,Li09,Garbarino09} reported that $T_c$ reaches 36$\sim$37 K at
higher pressure~\cite{Braithwaite09}, clearly putting FeSe$_{1-x}$ in the high $T_c$
category with iron pnictides having similar band filling~\cite{Subedi08}. The $T_c$ has also been increased by partial substitutions of the Se site with S or Te~\cite{Yeh08}. In particular, there are a number of reports on
FeSe$_{1-x}$Te$_x$~\cite{Taen09,Yadav09}, because crystal growth is easy and the material
is less toxic than Se or As.

The FeSe system is one of the most attractive systems for solving the mechanism of high $T_c$ superconductors,
because it has the simplest crystal structure among the iron-based superconductors reported.
In addition, the parent compound in FeSe exhibits the superconductivity\cite{Imai09} unlike other iron-based superconductors.
After reported Hsu \textit{et al.}, it has been reported that the synthesis method of the FeSe polycrystalline samples
has used mostly solid-state reaction method~\cite{Mizuguchi08,Garbarino09,Pomjakushina09,McQueen7909}. There are a few reports of the synthesis method of single crystalline samples
using a flux method~\cite{Braithwaite09,Wu09,Mok09,Zhang09} or vapor self-transport
method~\cite{Patel09}.  The above mentioned samples were synthesized with a Se deficiency for a nominal
composition of FeSe$_{1-x}$, and indicated  $T_c$$\sim$8.5 K or lower at ambient pressure~\cite{Margadonna08}. In addition, FeSe$_{1-x}$ with Se deficiency was reported to exhibit SDW transition/magnetic anomalies near 100 K~\cite{Sidorov09,Garbarino09,Zhang09}. On the other hand, Ref~\cite{McQueen7909} reported that FeSe exhibited superconductivity at a narrow range of stoichiometric Fe$_{1.01\pm0.02}$Se, or equivalently, FeSe$_{0.99\pm0.02}$ without magnetic ordering. This report is consistent with the NMR experimental results for stoichiometric FeSe compounds~\cite{Kotegawa08,Imai09}.
These suggest that magnetism is driven by anion vacancies in FeSe$_{1-x}$~\cite{Lee08}.

Our sample was synthesized using the Se self-flux method with a nominal compositions
of  Fe:Se=1:1.15 based on the phase diagram of FeSe, and the $T_c$ of our polycrystalline sample was 9.4 K~(zero resistivity) and 12.5 K (onset). In addition, EPMA revealed a Fe:Se composition ratio of 1:1~($\pm$0.02).
As like high $T_c$ cuprate, pseudogap behaviors appeared in the iron based superconductor 1111 system. Recently, photoemission spectroscopy(PES) of the normal state of FeSe revealed pseudogap-like behaviors.~\cite{Yamasaki09,Yoshida09}.

This paper reports that the FeSe superconductor exhibits pseudogap-like behavior near 110 K through an analysis of the resistivity using the resistivity curvature mapping (RCM) for high $T_c$ cuprates~\cite{Ando04}, the result of Seebeck coefficient measurement, and an analysis of the resistivity under applied magnetic field. Moreover, we report the data of critical current density of FeSe polycrystalline.

\section{Experimental}
\begin{figure}
\includegraphics{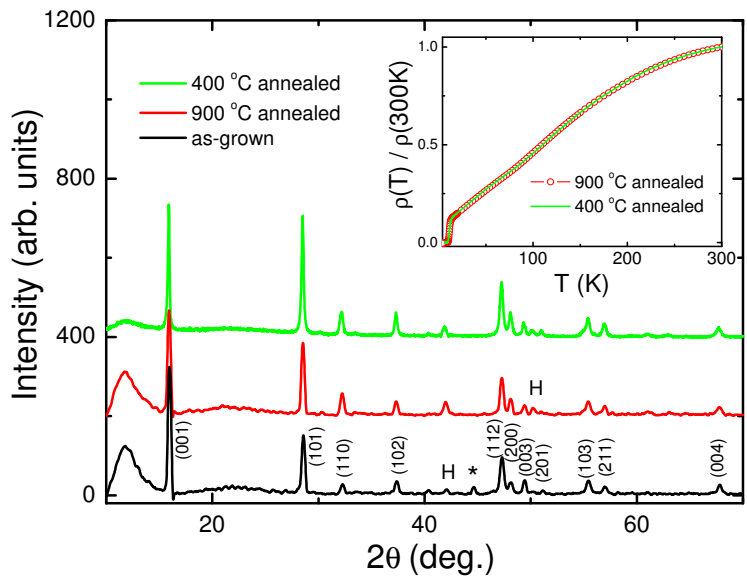}\centering
\caption{\label{fig.xrd} (color online) X-ray diffraction patterns of the as-grown, 900 $^o$C, and 400 $^o$C annealed FeSe sample.  The patterns show that the sample is composed of tetragonal FeSe ($P4/nmm$), with some hexagonal $\beta$-FeSe and Fe-phase as impurities. After annealing, the Fe-phase disappeared. As shown in the inset, the normalized resistivity of 400 $^o$C annealed sample is consistent with that of 900 $^o$C for full temperature range. }
\end{figure}

~~~~Polycrystalline samples with a nominal composition of Fe:Se=1:1.5 were prepared with
Fe powder (99.9$\%$) and Se grains (99.999$\%$). About 6.5 g of a stoichiometric quantity
was loaded into a small carbon crucible, which was then placed into a Mo crucible to avoid a reaction between Fe and Mo.
The cap covering the Mo crucible was welded by an arc welder filled Ar gas to prevent the escape of volatile Se. This arc-welded
Mo crucible was heated up to 1200 $^{\circ}$C for 8 hours and held at that temperature
for 122 hours in a self-fabricated vacuum furnace with tungsten mesh heater(VFTMH).
The Mo crucible was come down slowly from the heat source of the furnace at 1.5 mm/hour.
The obtained sample (as-grown sample) was polycrystalline with large grain boundaries including a small amount of a secondary phase.
The as-grown sample was then sealed into a small quartz tube under vacuum and annealed
at 900 $^{\circ}$C for 3 days. Powder XRD measurements of the as-grown and annealed sample
were carried out using a X-ray diffractometer (RIGAKU 12K) with Cu-K$\alpha$ radiation
from 2$\theta$$=$10$^\circ$ to 70$^\circ$ at a scanning rate of 0.02$^\circ$ per second.
The resistivity measurements were carried out in a Quantum Design PPMS system using the standard four-probe method from 4 K to room temperature at zero field, and from 4 to 200 K under a magnetic field of 1, 3, 5, 7, and 9 T. A Quantum Design MPMS was used to measure the temperature dependence of magnetization $M(T)$ at $H$=50 Oe, and the magnetic field dependence of magnetization $M(H)$ from -70 to 70 kOe. The Seebeck coefficient~(thermoelectric~power) was measured from 4 to 300 K under no magnetic field.

\section{Results and Discussions}

\subsection{XRD analysis}

~~~~Fig. 1 shows the X-ray diffraction pattern for as-grown sample, and the samples annealed at 900  $^o$C and 400  $^o$C. All the peaks were well indexed using the $P4/nmm$ space group from a Rietveld refinement. The calculated lattice constants
were $\textit{a}$= 3.7788 and $\textit{c}$=5.5310 ${\AA}$. Fe-phase and hexagonal $\beta$-FeSe were observed a bit in the as-grown sample. After annealing, the Fe-phase disappeared but $\beta$-FeSe remained as a secondary phase. Most recent studies annealed their samples at between 300$\sim$450 $^o$C based on Ref.~\cite{McQueen7909}. Therefore, the samples in this study were annealed at 400 $^o$C for 3~days. As shown Fig.1, the XRD patterns of the samples annealed at 400 $^o$C and 900 $^o$Cwere identical. The normalized resistivity of sample annealed at 400 $^o$C is consistent with that at 900 $^o$C for full temperature ranges as shown the inset in Fig.1. Hence, it implies that annealing temperature does not have sensitive effect on the $T_c$ and resistivity, unlike that reported in Ref.~\cite{McQueen7909}.

\subsection{Resistivity analysis}
\begin{figure}
\includegraphics{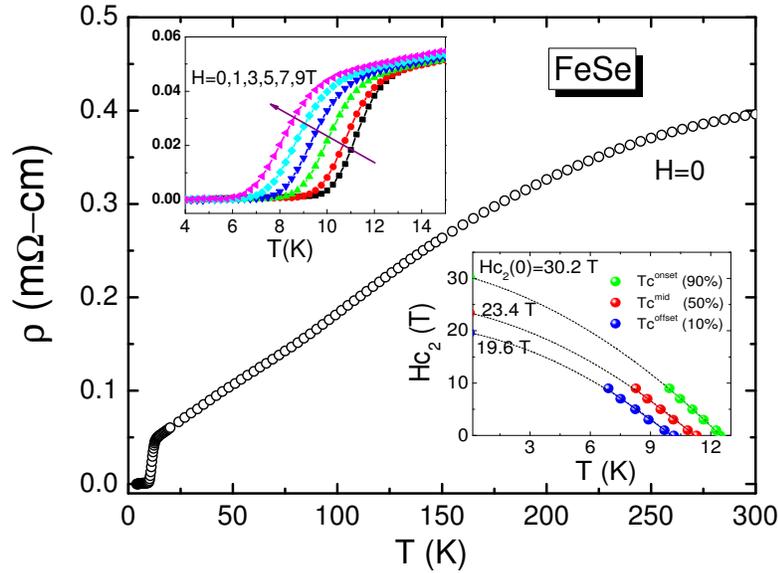}\centering
\caption{\label{fig:RT}(color online) Temperature dependence of the resistivity of FeSe. The left inset shows the resistivity measurements in magnetic fields of 0, 1, 3, 5, 7, and 9 T. The right inset shows the temperature dependence of the upper critical field $H_{c2}$ for $T_c^{onset}$, $T_c^{mid}$, and $T_c^{offset}$.}
\end{figure}

~~~~Fig. 2 shows the temperature dependence of the resistivity of the annealed sample at 900$^o$C from 4 K to room temperature. The resistivity began to decrease abruptly at $T_c^{onset}$=12.5 K due to superconductivity, and dropped to zero at $T_c^{zero}$=9.4 K. The $T_c^{zero}$ of our sample was $\sim$1 K higher than the 8$\sim$8.5 K reported previously. The ratio of room temperature to residual resistivity (RRR) was approximately 9.25. Furthermore, normal state residual resistivity value and resistivity value at 300 K are smaller than those of other polycrystallines~\cite{Mizuguchi08,Garbarino09,Pomjakushina09,McQueen7909}. These values are similar to recently reported data for single crystal FeSe~\cite{Braithwaite09}. Therefore, our polycrystalline sample is of good quality. The left inset in Fig. 2 shows the resistivity under a range of magnetic fields. With increasing magnetic field, the superconducting transitions shifted monotonously to lower temperatures. The right inset in Fig. 2 shows that the upper critical fields $H_{c2}(0)$ estimated using the Werthamer-Helfand-Hohenberg (WHH) formula $H_{c2}(0)$=-0.693$(dH_{c2}(T)/dT)$$\mid$$_{T_c}T_c$ for $T_c^{offset}$ (=$10\%\rho_n$, $\rho_n$ is the normal state resistivity value), $T_c^{mid}$ (=$50\%\rho_n$), and $T_c^{onset}$ (=$90\%\rho_n$) are 19.6, 23.4, and 30.2 T. These values are rather large considering that $T_c$ of FeSe occur at a low temperature. The coherence length $\xi$=3.3 nm was also estimated by using the Ginzberg Landau formula $H_{c2}$=$\Phi_0/2\pi\xi^2$, and was similar to that of a single crystal~\cite{Zhang09}.
\begin{figure}
\includegraphics{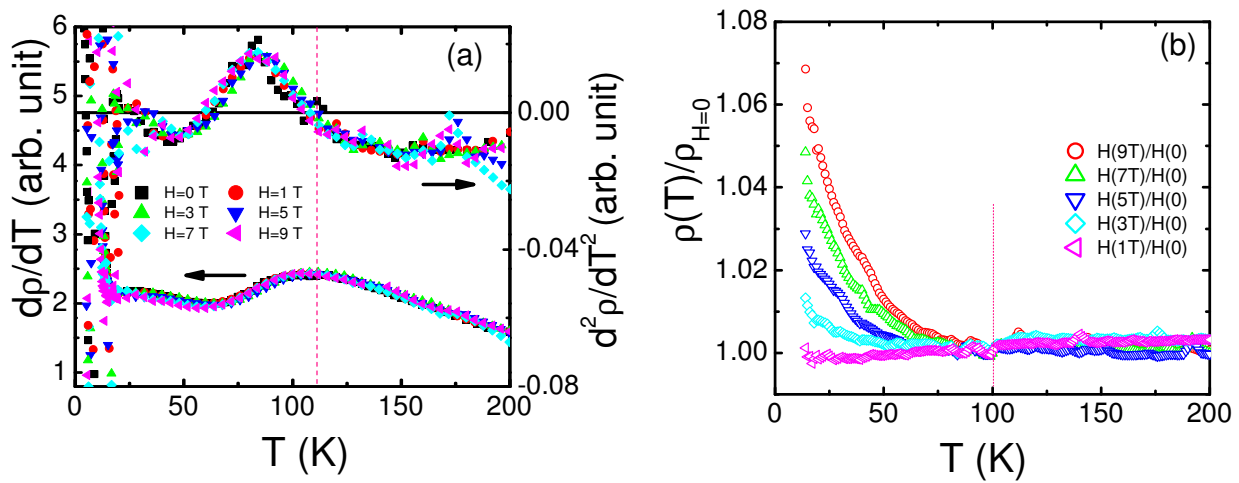}\centering
\caption{\label{fig:pseudogap}(color online) (a) In the graphs of  resistivity, $d\rho/dT$ and $d^2\rho/dT^2$ all curves showed anomaly near 110 K, i.e. characteristic temperature of a pseudogap $T^*$. The sign of  $d^2\rho/dT^2$  changes twice at $T^*$=110 K and 30 K. (b) Temperature dependence of $\rho_H(T)/\rho_{H=0}$ of FeSe under magnetic field. Increments of all curves on zero field resistivity increased below 100 K with decreasing temperature and increasing applied magnetic field. This provide evidence of an opening pseudogap near 100 K .}
\end{figure}

The resistivity curve of Fig. 2 showed a slight 'S' shape above $T_c$ and an inflection point near 110 K. This behavior is similar to that reported previously for a structural phase transition near 100 K for FeSe compounds ~\cite{Pomjakushina09,McQueen10309}. The 'S' shape of the resistivity curve is quite interesting because it was reported that an underdoped material of high $T_c$ superconducting cuprates displays this shape in $\rho(T)$~\cite{Timusk99}.  Ando $et$ $al.$ reported that the sign of $d^2\rho(T)/dT^2$ changes from positive to negative at the characteristic temperature of a pseudogap $T^*$ for high $T_c$ cuprates~\cite{Ando99}. The $d^2\rho(T)/dT^2$ and $d\rho(T)/dT$ of our sample shown in Fig. 3(a) revealed that the all curves are similar to the results of resistivity curvature mapping (RCM) reported by Ando $et$ $al$~\cite{Ando04}. As shown Fig 3.(a), signs of all $d^2\rho(T)/dT^2$  changed from negative to positive near 110 K.
It is different from Ando \textit{et al.}, that the sign of $d^2\rho(T)/dT^2$ changes again to
a positive sign near 30 K. This anomalous behavior is similar to the results
of the Seebeck coefficient mentioned below. More experimental and theoretical studies will be needed
 to make out the origin of this behavior near 30 K.
 Fig. 3(b) shows the temperature dependence of $\rho_H(T)/\rho_{H=0}$ from 4 to 200 K under
a magnetic field of 1, 3, 5, 7, and 9 T. As shown in Fig. 3(b), the increments of all curves on zero field resistivity increased below 100 K with decreasing temperature and increasing applied magnetic field~(maximum increment about 7 percent in 9 T above $T_c$). This temperature is similar to the analysis of the RCM method above and the enhancement of the antiferromagnetic spin fluctuation in the NMR experiment~\cite{Imai09}.

\subsection{Seebeck coefficient (thermoelectric~power) }%
\begin{figure}
\includegraphics{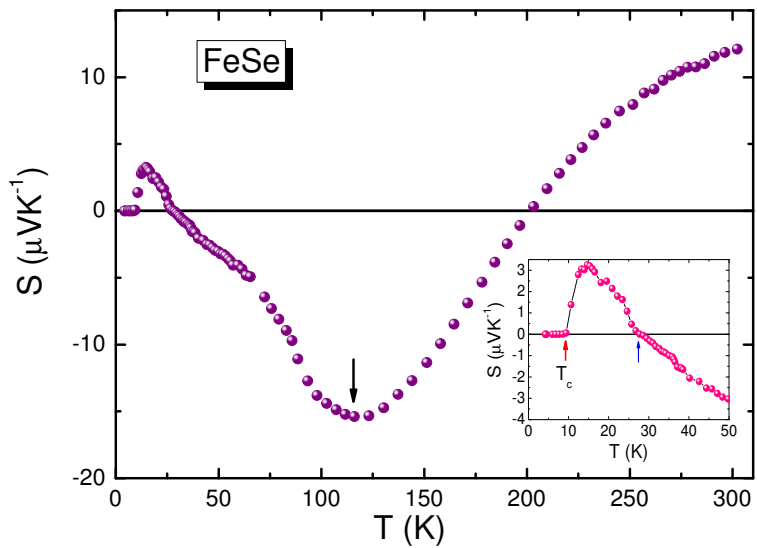}\centering
\caption{\label{fig:seebeck}Temperature dependence of the Seebeck coefficient of FeSe. The sign of the Seebeck coefficient of FeSe changes at approximately 200 K, and in the slope changes around 115 K. As shown the inset, the sign of S changed again near 30 K.}
\end{figure}
~~~~Fig. 4 shows the Seebeck coefficient measurements from 4 K to room temperature for FeSe. $S(T)$ is small and positive at room temperature. It then changes sign near 200 K~\cite{McQueen7909}, and passes through a negative maximum near 115 K, almost corresponding to the inflection temperature of the resistivity. The change in sign implies that electrons and holes make an almost equal contribution  to the conduction, as in a two carriers material~\cite{Wu07,Wu08}. Another interesting anomaly is that the sign of $S(T)$ changes again to positive near 30 K, which is consistent with the sign change of $d^2\rho(T)/dT^2$ near 30 K. There are no reports of a change in the sign of the Seebeck coefficient at low temperatures above the $T_c$. It is believed that this behavior is strongly associated with the mechanism of superconductivity.

However, it is unclear why the slope of the Seebeck coefficient changes near $T^*$,
in a similar manner to that observed with the resistivity. Previous reports of other materials suggest that the decrease in $\mid$$S(T)$$\mid$
above $T^*$ is due to an increasing number of thermally excited carriers across
the pseudogap~\cite{Nishino01,Nishino06,Lue02}. In addition, the results of several
experiments for iron based superconductors appear to suggest pseudogap-like behavior
~\cite{Yoshida09,Ishida08,Ichimura08,Sato08}.

\subsection{Magnetism}
\begin{figure}
\includegraphics{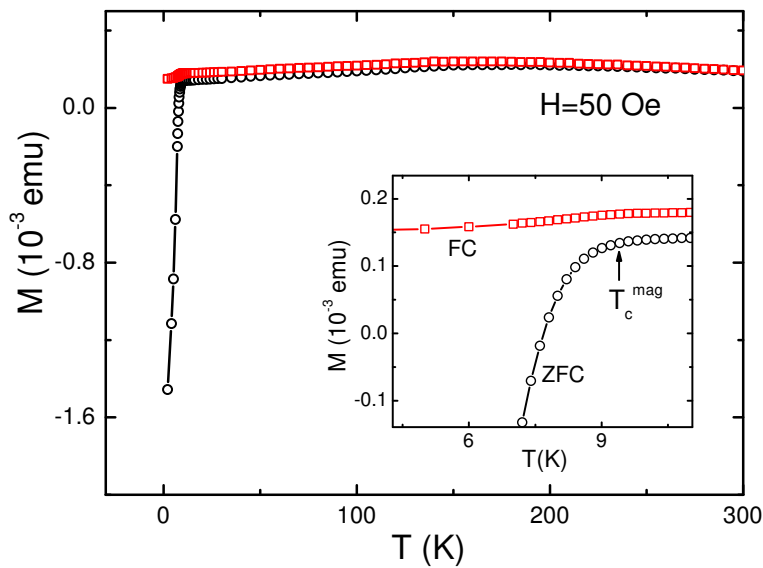}\centering
\caption{\label{fig:MT}(color online) Temperature dependence of the magnetization of FeSe measured in a 50 Oe. In zero field cool (ZFC), the magnetic onset of superconductivity appears at 9.4 K, which is the same as the zero resistance temperature.}
\end{figure}
~~~~Fig. 5 shows the temperature dependence of magnetization on zero-field cooling (ZFC) and field cooling (FC). The inset in Fig. 5 shows an enlargement of the low temperature range. A superconducting transition was observed at $T_c^{mag}$=9.4 K in ZFC, which is the same as the $T_c$ shown in Fig. 2. There was no magnetic anomaly from above $T_c$ to 300 K, which is similar to previous reports for a stoichiometric FeSe sample with a composition ratio Fe:Se=1:1.
The magnetization hysteresis loop in the inset of Fig. 6 shows a several isothermal $M$-$H$ curves at 2$\sim$9 K, which are eliminated ferromagnetic elements due to a secondary phase. The critical current density $J_c$ can be obtained from the $M$-$H$ loop using the Bean model. According to the Bean model, $J_c$ is given by $J_c$=20$\it{\Delta} M/[a(1-a/3b)]$, where $\it{\Delta} M$ is $M_{down}$-$M_{up}$, $M_{up}$ and $M_{down}$ are the magnetization on the sweeping fields up and down respectively, and $a$ and $b$ are the sample widths($a<b$). Fig. 6 shows the critical current density $J_c$ at several temperatures as a function of the field, and the critical current density calculated from the $M$-$H$ curves were estimated to be $10^2$$\sim$$10^4$ A/cm$^2$ at zero field. As shown Fig. 6, the critical current densities is $\sim$ 10$^3$-10$^4$ A/cm$^2$ uniformly up to 300 Oe in low temperature. The small value of $J_c$ in FeSe which is considerably smaller than those $10^5$$\sim$$10^6$ A/cm$^2$ of other iron-based superconductor samples~\cite{Taen09,Yadav09,Yang08} imply that our sample is good qualitative with few impurity. This difference might be due to the large grain boundaries of the sample. In addition, that may be ascribe to the low carrier and undoped compound.

\begin{figure}
\includegraphics{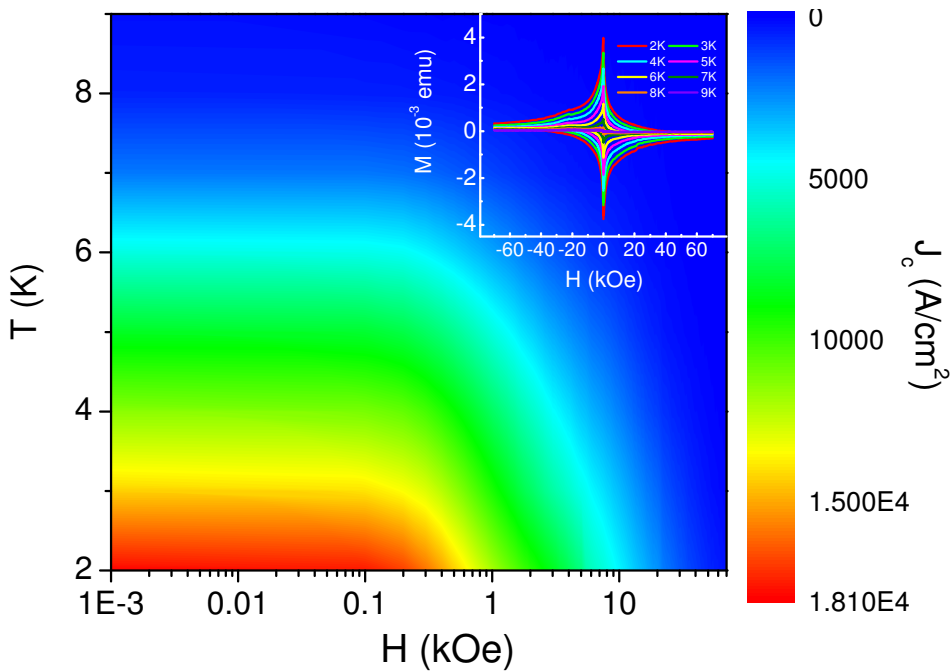}\centering
\caption{\label{fig:MH&Jc}(color online) (a) A typical superconducting magnetic hysteresis curve could be obtained by subtracting the $M$-$H$ curve at 11 K from that at temperatures below $T_c$.
(b) The diagram shows field dependence of the calculated critical current density for the temperature from the Bean's model. $J_c(H=0)$ was estimated to be $10^2$$\sim$$10^4$ A/cm$^2$.}
\end{figure}

\section{Conclusion}

~~~~A stoichiometric FeSe superconductor was synthesized with $T_c^{zero}$=9.4 K under a Se self-flux. The critical current density was measured to be $J_c$=$10^2$$\sim$$10^4$ A/cm$^2$ from an analysis of the $M$-$H$ loops. The Seebeck coefficient was 12.3 $\mu$V/K at room temperature, and the sign of $S(T)$ changed near 200 K as a two carriers material. In the resistivity measurements, the 'S' shape of the $\rho(T)$ curve, $d\rho^(T)/dT$, and $d^2\rho^(T)/dT^2$ exhibited pseudogap-like behavior at $T^*$=110 K by using the RCM method for high $T_c$ cuprates. In addition, the $\rho_H(T)/\rho_{H=0}$ under magnetic fields and the Seebeck coefficient $S(T)$ showed pseudogap-like behavior near $T^*$, nearly same pseudogap temperature in RCM. The other interesting and peculiar behavior is that the signs of $d^2\rho^(T)/dT^2$ and $S(T)$ changed from negative to positive near 30 K.

\ack{}
~~~~This study was performed for the Nuclear R$\&$D Programs funded
by the Ministry of Science $\&$ Technology (MOST) of Korea and by the Korea Research Foundation Grant funded by the Korean Government (KRF-2008-313-C00293)


\section*{References}

\begin{thebibliography}{35}

\bibitem{Hsu08} Fong-Chi Hsu, Jiu-Yong Luo, Kuo-Wei Yeh, Ta-Kun Chen, Tzu-Wen Huang, Phillip M. Wu, Yong-Chi Lee, Yi-Lin Huang, Yan-Yi Chu, Der-Chung Yan, and Maw-Kuen Wu, Proc. Natl. Acad. Sci. U.S.A. {\bf105}, 14262 (2008)

\bibitem{Mizuguchi08} Yoshikazu Mizuguchi, Fumiaki Tomioka, Shunsuke Tsuda, Takahide Yamaguchi and Yoshihiko Takano, Appl. Phys. Lett. {\bf93}, 152505 (2008)

\bibitem{Sidorov09} V.A. Sidorov, A.V. Tsvyashchenko, R.A. Sadykov, J. Phys : Condens Matter {\bf21}, 415701 (2009)

\bibitem{Li09} L. Li, Z.R. Yang, M. Ge, L. Pi, J.T. Xu, B.S. Wang, Y.P. Sun, Y.H. Zhang, J. Supercond. Nov. Magn. {\bf22}, 667-670 (2009)

\bibitem{Garbarino09} G. Garbarino, A. Sow, P. Lejay, A. Sulpice, P. Toulemonde, M. Mezouar and M. Nuez-Regueiro, EPL, {\bf86}, 27001 (2009)

\bibitem{Braithwaite09} D Braithwaite, B Salce, G Lapertot, F Bourdarot, C Marin, D Aoki and M Hanfland, Superconducting and normal phases of FeSe single crystals at high pressure, J. Phys:Condens Matter {\bf21}, 232202(2009)

\bibitem{Subedi08} Alaska Subedi, Lijun Zhang, D.J. Singh, and M.H. Du, Phys. Rev. B {\bf78}, 134514 (2008)

\bibitem{Kotegawa08} Hisashi Kotegawa, Satoru Masaki, Yoshiki Awai, Hideki Tou, Yoshikazu Mizuguchi, and Yoshihiko Takano, J. Phys. Soc. Jpn. {\bf77},113703 (2008)

\bibitem{Imai09} T. Imai, K. Ahilan, F. L. Ning, T. M. McQueen, and R. J. Cava, Phys. Rev. Lett. {\bf102}, 177005 (2009)

\bibitem{Lee08} K.-W. Lee, V. Pardo, and W. E. Pickett, Phys. Rev. B {\bf78}, 174502 (2008)

\bibitem{Pomjakushina09} E. Pomjakushina, K. Conder, V. Pomjakushin, M. Bendele, and R. Khasanov, Phys. Rev. B {\bf80}, 024517 (2009)

\bibitem{McQueen7909} T. M. McQueen, Q. Huang, V. Ksenofontov, C. Felser, Q. Xu, H. Zandbergen, Y. S. Hor, J. Allred, A. J. Williams, D. Qu, J. Checkelsky, N. P. Ong, and R. J. Cava, Phys. Rev. B {\bf79}, 014522 (2009)

\bibitem{Yeh08} K.W. Yeh, H.C. Hsu, T.W. Haung, P.M. Wu, Y.L. Huang, T.K. Chen, J.Y. Luo and M.K. Wu, J. Phys. Soc. Jpn. {\bf77}, Suppl. C, pp.19-22 (2008)

\bibitem{Taen09} T. Taen, Y. Tsuchiya, Y. Nakajima, and T. Tamegai, Phys. Rev. B {\bf80}, 092502 (2009)

\bibitem{Yadav09} C. S. Yadav and P. L. Paulose, New J. Phys. {\bf11} 103046 (2009)

\bibitem{Wu09} M.K. Wu, F.C. Hsu, K.W. Yeh, T.W. Huang, J.Y. Luo, M.J. Wang, H.H. Chang, T.K. Chen, S.M. Rao ,B.H. Mok, C.L. Chen, Y.L. Huang, C.T. Ke, P.M. Wu, A.M. Chang, C.T. Wu, T.P. Perng , Physica C {\bf469}, 340-349 (2009)

\bibitem{Mok09} B. H. Mok, S. M. Rao, M. C. Ling, K. J. Wang, C. T. Ke, P. M. Wu, C. L. Chen, F. C. Hsu, T. W. Huang, J. Y. Luo, D. C. Yan, K. W. Ye, T. B. Wu, A. M. Chang, and M. K. Wu, CRYSTAL GROWTH $\&$ DESIGN  {\bf9},NO.7, 3260-3264 (2009)

\bibitem{Zhang09} S. B. Zhang, Y. P. Sun, X. D. Zhu, X. B. Zhu, B. S. Wang, G. Li, H. C. Lei, X. Luo, Z. R. Yang, W. H. Song and J. M. Dai, Supercond. Sci. Technol. {\bf22},015020 (2009)

\bibitem{Patel09} U. Patel, J. Hua, S. H. Yu, S. Avci, Z. L. Xiao, H. Claus, J. Schlueter, V. V. Vlasko-Vlasov, U. Welp, and W. K. Kwok, Appl. Phys. Lett. {\bf94}, 082508 (2009)

\bibitem{Margadonna08} Serena Margadonna, Yasuhiro Takabayashi, Martin T. McDonald, Karolina Kasperkiewicz, Yoshikazu Mizuguchi, Yoshihiko Takano, Andrew N. Fitch, Emmanuelle Suard and Kosmas Prassides, Chem. Commun. 5607 (2008)

\bibitem{McQueen10309} T. M. McQueen, A. J. Williams, P. W. Stephens, J. Tao, Y. Zhu, V. Ksenofontov, F. Casper, C. Felser, and R. J. Cava, Phys. Rev. Lett. {\bf103}, 057002 (2009)

\bibitem{Timusk99} Tom Timusk and Bryan Statt, Rep. Prog. Phys. {\bf62},61-62 (1999)

\bibitem{Ando99} Yoichi Ando and T. Murayama, Phys. Rev. B {\bf60}, NUMBER 10 (1999)

\bibitem{Ando04} Yoichi Ando, Seiki Komiya, Kouji Segawa, S. Ono, and Y. Kurita, Phys. Rev. Lett. {\bf93}, 267001 (2004)

\bibitem{Yang08} Huan Yang, Huiqian Luo, Zhaosheng Wang, and Hai-Hu Wen, Appl. Phys. Lett. {\bf93}, 142506 (2008)

\bibitem{Wu07} X. J. Wu, D. Z. Shen, Z. Z. Zhang, J. Y. Zhang,aK. W. Liu, B. H. Li, Y. M. Lu, B. Yao, D. X. Zhao, B. S. Li, C. X. Shan, and X. W. Fan, H. J. Liu, C. L. Yang, Appl. Phys. Lett. {\bf90}, 112105 (2007)

\bibitem{Wu08} X. J. Wu, Z. Z. Zhang, J. Y. Zhang, B. H. Li, Z. G. Ju, Y. M. Lu, B. S. Li, and D. Z. Shen, J. Appl. Phys. {\bf103}, 113501 (2008)

\bibitem{Nishino01} Y. Nishino, H. Kato, and M. Kato, Phys. Rev. B {\bf63}, 23303 (2001)

\bibitem{Nishino06} Y. Nishino and S. Deguchi, Phys. Rev. B {\bf74}, 115115 (2006)

\bibitem{Lue02} C. S. Lue and Y.K Kuo, Phys. Rev. B {\bf66}, 085121 (2002)

\bibitem{Yamasaki09} A. Yamasaki, S. Imada, K. Takase, T. Muro, Y. Kato, H. Kobori, A. Sugimura, N. Umeyama, H. Sato, Y. Hara, N. Miyakawa, S. I. Ikeda, arXiv : 0902.3314

\bibitem{Yoshida09} Rikiya Yoshida, Takanori Wakita, Hiroyuki Okazaki, Yoshikazu Mizuguchi, Shunsuke Tsuda, Yoshihiko Takano, Hiroyuki Takeya, Kazuto Hirata, Takayuki Muro, Mario Okawa, Kyoko Ishizaka, Shik Shin, Hisatomo Harima, Masaaki Hirai, Yuji Muraoka, and Takayoshi Yokoya, J. Phys. Soc. Jpn. {\bf78}, 034708 (2009)

\bibitem{Ishida08} Y. Ishida, T. Shimojima, K. Ishizaka, T. Kiss, M. Okawa, T. Togashi, S. Watanabe, X.-Y. Wang, C.-T. Chen, Y. Kamihara, M. Hirano, H. Hosono, and S. Shin, J. Phys. Soc. Jpn. {\bf77}, Suppl. C, pp. 61-64 (2008)

\bibitem{Ichimura08} Koichi Ichimura, Junya Ishioka, Toru Kurosawa, Katsuhiko Inagaki, Migaku Oda, Satoshi Tanda, Hiroki  Takahashi, Hironari Okada, Yoichi Kamihara, Masahiro Hirano, and Hideo Hosono, J. Phys. Soc. Jpn, {\bf77}, 151     (2008)

\bibitem{Sato08} Takafumi Sato, Seigo Souma, Kosuke Nakayama, Kensei Terashima, and Hideo Hosono, J. Phys. Soc. Jpn,     {\bf77}, 063708 (2008)


\end{thebibliography}
\end{document}